\documentclass[twocolumn,showpacs,amssymb]{revtex4}
\usepackage{graphicx}% Include figure files

\begin{document}

\title{Quantum coherence of double-well BEC: a SU(2)-coherent-state path-integral
approach}
\author{Yi Zhou, Hui Zhai, Rong L\"{u}, Zhan Xu, Lee Chang}
\address{Center for Advanced Study, Tsinghua University, Beijing 100084, P. R. China}
\email{yizhou@castu.tsinghua.edu.cn}
\begin{abstract}
Macroscopic quantum coherence of Bose gas in a double-well potential is
studied based on SU(2)-coherent-state path-integral. The ground state 
and fluctuations around it can be obtained by this method. In this picture, one
can obtain macroscopic quantum superposition states for attractive Bose gas. The
coherent gap of degenerate ground states is obtained with the instanton
technique. The phenomenon of macroscopic quantum self-trapping is also
discussed.
\end{abstract}
\date{\today}
\maketitle

\section{ Introduction}

Quantum tunneling at mesoscopic scale is one of the most fascinating
phenomena in condensed matter physics. The double-well potential provides a
simple and yet physically relevant example for studies of quantum tunneling
in mesoscopic systems. Recently, there have been great experimental and
theoretical interests in studying the coherent quantum tunneling between two
Bose-Einstein condensates (BEC) in a double-well potential. It was found
that the ground state changes from a coherent state to a Fock state as the
interaction between particles is increased in a repulsive Bose gas\cite
{Rokshar,Dalibard}. It was numerically shown that the attractive Bose gas in
a double-well potential has Schr\"{o}dinger Cat-like ground states\cite{TLH}. 
A similar model can be found in Refs.\cite{MiIburn}, \cite{Vardi} and \cite
{Parkins_Walls}.

In this paper we study quantum coherence of BEC in a double-well potential
by mapping the two site boson model onto an anisotropic spin model in an
external magnetic field\cite{MiIburn}. Then the coherence properties of Bose
system can be studied in the SU(2)-coherent-state path-integral
representation. From the effective classical energy, it is easy to show that
the phenomenon of macroscopic quantum self-trapping (MQST) exists in both
the repulsive and attractive interaction cases. The point separating the
coherent ground state and macroscopic quantum superposition state can be obtained
analytically. The number fluctuation and relative phase fluctuation between
two-well condensates are given through path-integral technique. Within the
instanton technique, we obtain the tunnel splitting of degenerate ground
states in the attractive interaction case. It is noted that the model
presented here is general (see Eq.(\ref{H})), including BEC in
symmetric or non-symmetric double-well potential,
and in different regimes of
parameters. We emphasize that the quantum coherence properties of BEC depend
on the parameters of system distinctly.

\section{Physical model and SU(2) coherent state}
{\it \ }For the Bose gas in an external potential, $U\left( \vec{r}\right) $%
, the Hamiltonian can be written in the second quantized form as 
\begin{equation}
H=\int d\vec{r}\psi ^{\dag }\left( -\frac{\hbar ^2}{2m}\nabla ^2+U\right)
\psi +\frac g2\int d\vec{r}\psi ^{\dag }\psi \psi ^{\dag }\psi .  \label{H1}
\end{equation}
In Eq. (\ref{H1}) we have used a shape-independent form for the atom-atom
interaction with $g=4\pi \hbar ^2a_{sc}/m$, where $a_{sc}$ is the s-wave
scattering length for repulsive ($g>0$) or attractive ($g<0$) interactions.
Under the two-mode approximation\cite{Parkins_Walls}, $\psi $ can be
expanded as 
\[
\psi =\phi _Lb_L+\phi _Rb_R, 
\]
where $\phi _L$ and $\phi _R$ are real, and describe the mainly left-well
and mainly right-well populated states respectively. With the help of
Schwinger boson representation for angular momentum\cite{MiIburn}, 
\begin{eqnarray}
J_{+} &\equiv &J_x+iJ_y=b_L^{\dag }b_R,J_{-}\equiv J_{+}^{\dag },J_z=\frac 12%
\left( b_L^{\dag }b_L-b_R^{\dag }b_R\right) ,  \nonumber \\
N &=&b_L^{\dag }b_L+b_R^{\dag }b_R,J^2=\frac N2\left( \frac N2+1\right) ,
\end{eqnarray}
we can map the Bose model (1) to a new Hamiltonian:

\begin{eqnarray}
H &=&\varepsilon N-2t^{\prime }J_x+\frac{\beta _L}2\left( \frac N2%
+J_z\right) ^2+\frac{\beta _R}2\left( \frac N2-J_z\right) ^2  \nonumber \\
&&+u\left( 2J_x^2-J_z^2+\frac{N^2}4\right) +v_L\left(
NJ_x+J_zJ_x+J_xJ_z\right)  \nonumber \\
&&+v_R\left( NJ_x-J_zJ_x-J_xJ_z\right) ,  \label{H}
\end{eqnarray}
where $\varepsilon =\int d\vec{r}\phi _{L(R)}\left( -\frac{\hbar ^2}{2m}%
\nabla ^2+U\right) \phi _{L(R)}$, $\beta _{L\left( R\right) }=g\int d\vec{r}%
\phi _{L\left( R\right) }^4$, $t^{\prime }=-\int d\vec{r}\phi _{L(R)}\left( -%
\frac{\hbar ^2}{2m}\nabla ^2+U\right) \phi _{R(L)}$, $u=g\int d\vec{r}\phi
_L^2\phi _R^2$, $v_{L\left( R\right) }=g\int d\vec{r}\phi _{L\left( R\right)
}^3\phi _{R\left( L\right) }$. One can easily estimate that $|\beta
_{L\left( R\right) }|\gg |u|,|v_{L\left( R\right) }|$. And the condition $%
t^{\prime }>0$ can always be satisfied by choosing proper signs (positive or negative)
for real $\phi_L$ and $\phi _R$. The Hamiltonian (\ref{H}) describes an anisotropic spin
system in an external magnetic field, which has been studied extensively in
quantum tunneling in mesoscopic magnets\cite{Garg}. The terms involving $N$
but independent of components of $J$ are constants and can be dropped in $H$.
The Hamiltonian Eq.(\ref{H}) is general, valid for the 
symmetric and non-symmetric wells as well. By non-symmetric well we mean
the difference in the depth and shape between left and right wells, which can be tuned
by the external potential $U\left( \vec{r}\right)$.

For simplicity, we consider the symmetrical-well case, $\beta _L=\beta
_R=\beta ^{\prime }$, $v_L=v_R=v$. Hence the Hamiltonian can be simplified
as $H=-2tJ_x+\beta J_z^2+2uJ_x^2$, where $\beta =\beta ^{\prime }+u$ and $%
t=t^{\prime }-vN$. In this paper, we consider only the case $|\beta |\gg |u|$
and $t>0$. In fact, the two-mode approximation is invalid for large $vN$
for the occupation of other modes\cite{Parkins_Walls}. The Hamiltonian now
reads:

\begin{equation}
H=-2tJ_x+\beta J_z^2.
\end{equation}
In the SU(2)-coherent-state path-integral representation, the Euclidean
transition amplitude from an initial state to a final state can be written as%
\cite{Auerbach} 
\begin{equation}
\left\langle \Omega _f\right| e^{-H(\tau _f-\tau _i)/\hbar }\left| \Omega
_i\right\rangle =\int \left[ d\Omega (\tau )\right] \exp \left[ -\frac 1\hbar
S_E(\theta ,\phi )\right] ,
\end{equation}
where 
\begin{equation}
S_E(\theta ,\phi )=\int_{\tau _i}^{\tau _f}d\tau \left[ i\hbar \frac N2%
(1-\cos \theta )\left( \frac{d\phi }{d\tau }\right) +E(\theta ,\phi )\right]
.  \label{Se}
\end{equation}
The first term in Eq. (\ref{Se}) is the Wess-Zumino term, and the effective
classical energy $E(\theta ,\phi )$ is 
\begin{equation}
E(\theta ,\phi )=-tN\sin \theta \cos \phi +\frac{\beta N^2}4\cos ^2\theta .
\end{equation}
It is noted that the action (\ref{Se}) describes the $\left( 1\oplus
1\right) $-dimensional dynamics in the Hamiltonian formulation, which
consists of the canonical coordinates $\phi $ and the canonical momentum $%
p_\phi =i\hbar N\left( 1-\cos \theta \right) /2$. In this picture, $\langle
J_x\rangle =\frac N2\sin \theta \cos \phi $, $\langle J_y\rangle =\frac N2%
\sin \theta \sin \phi $, $\langle J_z\rangle =\frac N2\cos \theta
=(N_L-N_R)/2$, where $N_L$ and $N_R$ are the left and right-well particle
number respectively. Therefore, $\cos \theta =(N_L-N_R)/N$ corresponds to
the double-well population imbalance, and $\langle b_L^{\dagger }b_R\rangle
=\langle J_{+}\rangle =\frac N2\sin \theta e^{i\phi }$, $\phi $ corresponds
to the phase difference between the double-well condensates.

\section{Effective classical energy}

In this section, we will show that the effective classical energy and the classical
equations of motion give many interesting results about quantum coherence
properties of BEC in a double-well potential. It is natural to investigate
the classical orbits on the Bloch sphere, which can be described by the
energy contour. In Figs. 1(a)-(d), we plot the classical orbits for
different parameters $\beta N/2t$. The classical orbits show the interesting
phenomenon of self-maintained population imbalance, i.e., macroscopic
quantum self-trapping. This phenomenon was first found by Smerzi et al. in
both repulsive and attractive interaction cases by using the canonical conjugate variables
approximation and numerical calculation, and was explained as a nonlinear
phenomenon induced by the interaction between atoms\cite{smerzi}. Here with
the help of effective classical energy, we study this MQST phenomenon
in both repulsive and attractive interaction cases, and present the
condition for MQST analytically. %
%One can see that the MQST phenomenon happen when the classical orbits are not symmetrical about the axis $\theta =\pi /2$ which correspond to the zero population imbalance.
From the effective classical energy $E(\theta ,\phi )$ one can easily obtain
the two dividing points $\beta N/2t=\pm 1$, which correspond to the
existence of two degenerate energy maxima at $\phi =\pi $ (i.e. the phase
difference between double-well BEC is $\pi $) or minima at $\phi =0$ (i.e.
the phase difference is $0$). When $-1<\beta N/2t<1$, Fig. 1 show that MQST
is forbidden and the particle number of each well oscillates around $N/2$.
However the phenomena of MQST exist in the cases of $\beta N/2t>1$ and $%
\beta N/2t<-1$. As a result, we conclude that the phenomenon of MQST is
permitted when the interaction between atoms is strong enough to obtain
degenerate energy maxima or minima ($|\beta |N/2t>1$) in both repulsive and
attractive interaction cases. Our results of repulsive case agree well with
the results in Refs.\cite{Vardi} and \cite{smerzi}.

\begin{figure}[tbp]
\begin{center}
\includegraphics[width=2.7in]%
{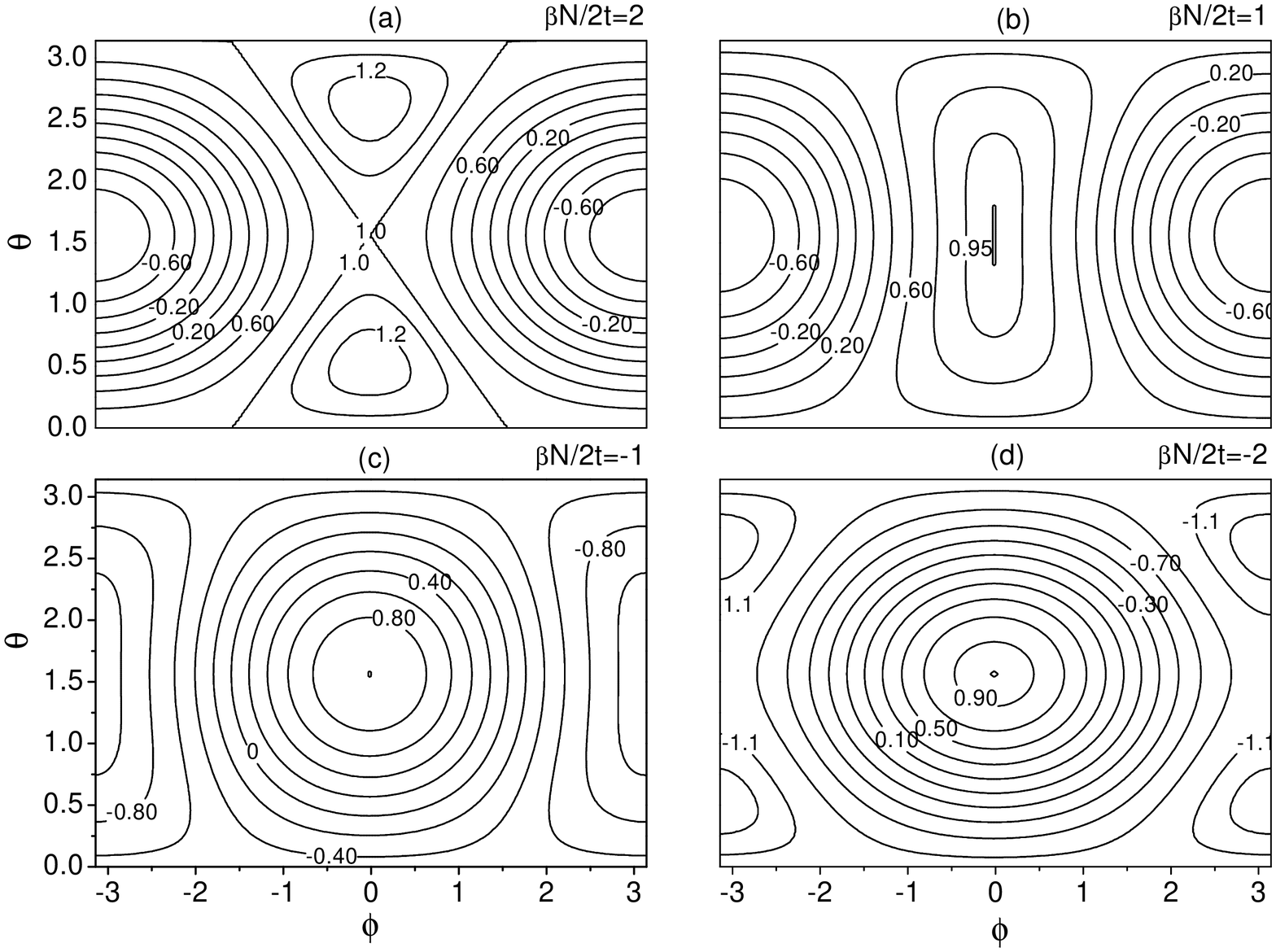}%
\caption{ Classical orbits (energy scale is $tN$) for different parameters $%
\beta N/2t$. $\cos \theta $ corresponds to the double-well population
imbalance and $\phi $ corresponds to the phase difference between the
double-well condensates. Degenerate energy maxima (or minima) imply the
phenomena of MQST. Figs. 1(a) and (b) are for the repulsive case, Figs. 1(c)
and (d) are for the attractive case. Two critical points correspond to $%
\beta N/2t=\pm 1$. }
\end{center}
\label{Fig.1}
\end{figure}

Another interesting observation concerns  macroscopic quantum superposition state
in BEC with attractive interaction. One can easily show that the system has
different energy minima for different parameters $\beta N/2t$. The energy
minima appear at $\phi =0$ and different $\theta $ (i.e. different
population imbalance). The $\theta $-dependence of effective classical
energy $E(\theta ,\phi =0)$ is plotted in Fig. 2 for different parameter $%
\beta N/2t$. When $\beta N/2t>-1$, there is only one energy minimum at $%
\theta =\pi /2$; while when $\beta N/2t<-1$, there are two degenerate energy
minima at $\theta =\theta _0$ and $\theta =\pi -\theta _0$, where $\sin
\theta _0=2t/|\beta |N$. The former case favors a coherent ground state or a
Fock ground state\cite{Rokshar,TLH}, the latter case favors a
Schr\"{o}dinger cat-like ground state which is the superposition of two SU(2)
coherent states $\left| \theta _0,0\right\rangle $ and $\left| \pi -\theta
_0,0\right\rangle $. This kind of definition describes the same
Schr\"{o}dinger cat-like state as that in Ref.\cite{TLH} , where the result
was obtained by numerical calculation. Here the dividing
point separating the coherent ground state and the Schr\"{o}dinger cat-like state
is analytically obtained in a simple and clear approach. Moreover, the
tunnel splitting can be obtained by applying the instanton technique in the
SU(2)-coherent-state path-integral representation, as shown in the next section.

\begin{figure}[tbp]
\begin{center}
\includegraphics[width=2.7in]%
{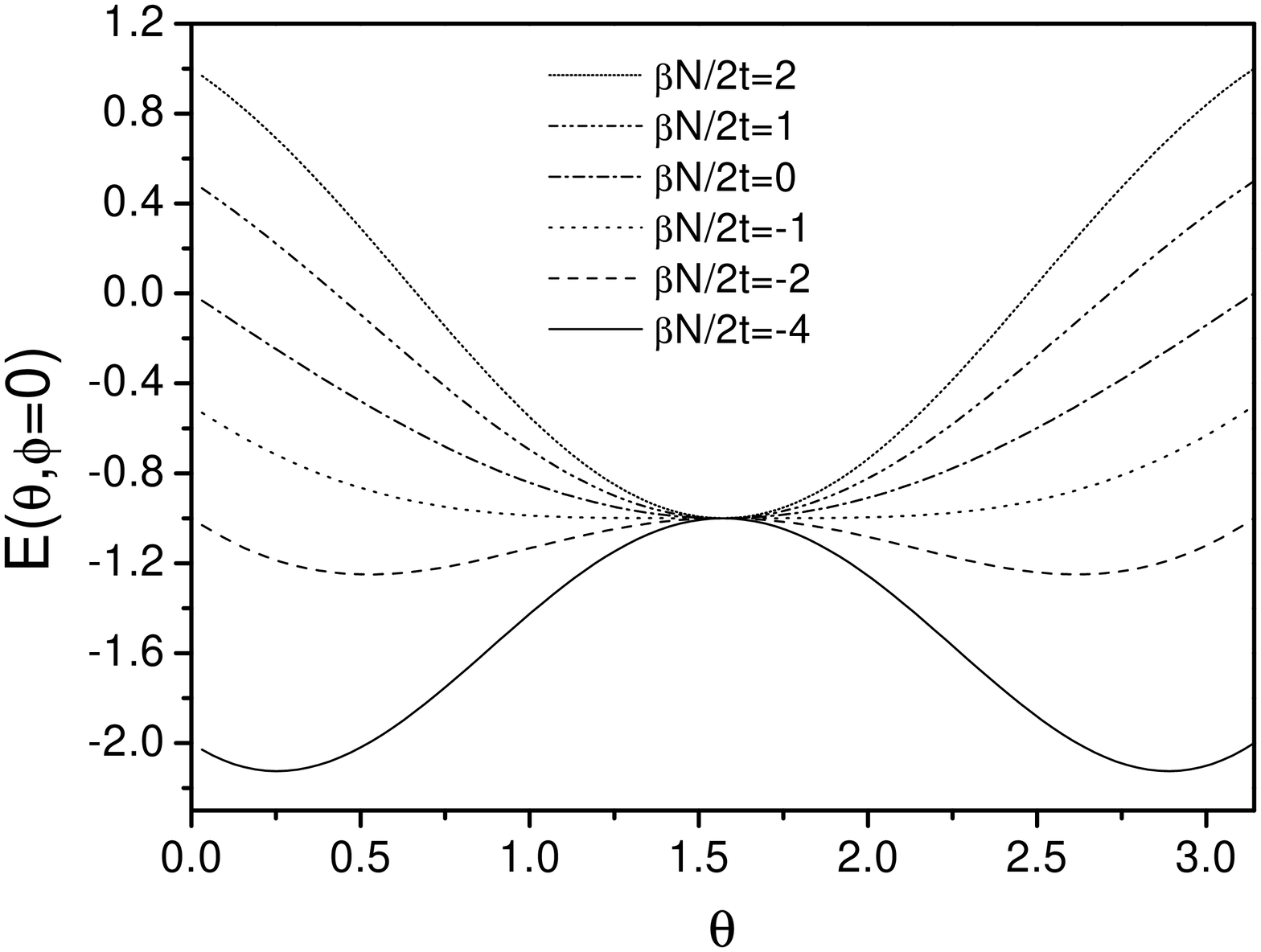}%
\caption{$\theta $-dependence of effective classical energy for different
parameters $\beta N/2t$. }
\end{center}
\label{Fig.2}
\end{figure}
\section{Coherent Quantum Tunneling}

As shown in Fig. 2, there will be two degenerate energy minima when $\beta
N/2t<-1$ for the attractive BEC. Now the system in question performs
coherent quantum tunneling (i.e., quantum coherence or coherent
superposition) between two degenerate energy minima. The tunneling removes
the degeneracy of the original ground states, and the true ground state
(i.e., Schr\"{o}dinger cat-like state) is a superposition of the previous ground
states. Tunneling between neighboring degenerate vacua can be described by
the instanton configuration and leads to a level splitting of the ground
states. Here we evaluate the tunnel splitting of two degenerate ground
states by applying the instanton technique\cite{Coleman}.

After adding some constants, we can rewrite the effective energy as

\begin{equation}
E(\theta ,\phi )=\frac{\left| \beta \right| N^2}4\left( \sin \theta -\sin
\theta _0\right) ^2+tN\sin \theta \left( 1-\cos \phi \right) ,  \label{E_eff}
\end{equation}
where $\sin \theta _0=2t/|\beta |N$. From $\delta S_E(\theta ,\phi )=0$, we
obtain the instanton solution as 
\begin{eqnarray}
\cos \overline{\theta } &=&-\cos \theta _0\tanh \left( \omega _b\tau \right)
,  \nonumber \\
\sin \overline{\phi } &=&-\frac i2\frac{\cot ^2\theta _0%
%TCIMACRO{\func{sech} }
%BeginExpansion
\mathop{\rm sech}%
%EndExpansion
^2\left( \omega _b\tau \right) }{\left[ 1+\cot ^2\theta _0%
%TCIMACRO{\func{sech} }
%BeginExpansion
\mathop{\rm sech}%
%EndExpansion
^2\left( \omega _b\tau \right) \right] ^{1/2}},
\end{eqnarray}
corresponding the transition from $\theta =\theta _0$ to $\theta =\pi
-\theta _0$, where $\omega _b=N|\beta |\cos \theta _0/2\hbar $. The
associated classical action is found to be 
\begin{equation}
S_{cl}=N\left[ -\cos \theta _0+\frac 12\ln \left( \frac{1+\cos \theta _0}{%
1-\cos \theta _0}\right) \right] ,  \label{Scl}
\end{equation}
and the final result of tunnel splitting is 
\begin{equation}
\hbar \Delta =8\left( \frac N{2\pi }\right) ^{1/2}|\beta |N\frac{\left( \cos
\theta _0\right) ^{5/2}}{\sin \theta _0}\left( \frac{1-\cos \theta _0}{%
1+\cos \theta _0}\right) ^{\frac 12\cos \theta _0}e^{-S_{cl}}.
\end{equation}

It is noted that the tunnel splitting is obtained with the help of instanton
technique in the SU(2)-coherent-state path-integral representation, which is
semiclassical in nature, i.e., valid for large $N$. Therefore, one should
analyze the validity of the semiclassical approximation. The semiclassical
approximation is valid only when the energy splitting is far less than the
energy barrier $N^2|\beta |\left( 1-\sin \theta _0\right) ^2/4$ and the
energy of zero point oscillation $N^2|\beta |\cos ^2\theta _0/4$, which
indicates that the classical action $S_{cl}\gg 1$. From Eq. (\ref{Scl}) one
can easily see $S_{cl}\gg 1$ when $\theta _0^{\prime }=$ $\pi /2-0.3$ for
typical particle number of attractive BEC $N=1000$. Then the semiclassical
approximation should be already rather good for $0\leq \theta _0\leq \theta
_0^{\prime }$.

Because of coherent quantum tunneling, the two degenerate energy minima
ground state $\left| \theta _0,0\right\rangle $ and $\left| \pi -\theta
_0,0\right\rangle $, which correspond to SU(2) coherent states with
population imbalance $\pm \cos \theta _0$ and phase difference $0$ between
condensates in the two wells, split to the two parity-different
Schr\"{o}dinger cat states $\left| \pm \right\rangle =\left( \left| \theta
_0,0\right\rangle \pm \left| \pi -\theta _0,0\right\rangle \right) /\sqrt{2}$
with energy splitting $\hbar \Delta $. One can show that the energy
splitting is extremely small when $\theta _0$ is away from $\pi /2$.

\section{Number fluctuations and relative phase fluctuations}

In this section we consider the fluctuations around the ground state. We
shall evaluate the fluctuations for the parameters $\theta $ and $\phi $,
which correspond to the relative number fluctuations and the phase
fluctuations. Rewriting the parameters as $\theta (\tau )=\bar{\theta}(\tau
)+\theta _1(\tau )$ and $\phi (\tau )=\bar{\phi}(\tau )+\phi _1(\tau )$, one
obtains the Euclidean action as 
\begin{equation}
S_E(\theta ,\phi )=S_{cl}+\frac 12\delta ^2S,
\end{equation}
where $S_{cl}$ is the classical action which satisfy $\delta S_{cl}=0$ and

\begin{eqnarray}
\frac 12\delta ^2S &=&-\int_{\tau _i}^{\tau _f}i\hbar \frac N2\frac d{d\tau }%
\left( \sin \bar{\theta}\theta _1\right) \phi _1d\tau  \nonumber \\
&&+\frac 12\int_{\tau _i}^{\tau _f}i\hbar \frac N2\cos \bar{\theta}\stackrel{%
.}{\bar{\phi}}\theta _1^2d\tau \\
&&+\frac 12\int_{\tau _i}^{\tau _f}\left( E_{\theta \theta }\theta
_1^2+2E_{\theta \phi }\theta _1\phi _1+E_{\phi \phi }\phi _1^2\right) d\tau .
\nonumber
\end{eqnarray}
In the above equation, $E_{\theta \theta }=\partial ^2E/\partial \theta ^2$, 
$E_{\theta \phi }=\partial ^2E/\partial \theta \partial \phi $ and $E_{\phi
\phi }=\partial ^2E/\partial \phi ^2$, which are evaluated at the classical
path. Under the condition that $E_{\phi \phi }>0$, the Gaussian integration
can be performed over $\phi _1$, then the effective action for $\theta _1$
is found to be

\begin{equation}
I\left( \theta _1\right) =\int_{\tau _i}^{\tau _f}\left( A\dot{\theta}%
_1^2+B\theta _1\dot{\theta}_1+C\theta _1^2\right) d\tau .
\end{equation}
Using Eq. (\ref{E_eff}), we have

\begin{eqnarray*}
E_{\phi \phi } &=&tN\sin \bar{\theta}\cos \bar{\phi}>0, \\
A &=&\frac{\hbar ^2N\sin \bar{\theta}}{8t\cos \bar{\phi}}, \\
B &=&0, \\
C &=&\frac{tN\cos \bar{\phi}}{2\sin \bar{\theta}}+\frac{\beta N^2}4\sin ^2%
\bar{\theta}.
\end{eqnarray*}
The effective action reads:

\begin{eqnarray}
I\left( \theta _1\right) &=&\int_{\tau _i}^{\tau _f}\left[ \frac{\hbar
^2N\sin \bar{\theta}}{8t\cos \bar{\phi}}\dot{\theta}_1^2\right.  \nonumber \\
&&+\left. \left( \frac{tN\cos \bar{\phi}}{2\sin \bar{\theta}}+\frac{\beta N^2%
}4\sin ^2\bar{\theta}\right) \theta _1^2\right] d\tau .
\end{eqnarray}

In the case of $\beta N/2t>-1$, the classical ground state is $\bar{\theta}%
=\pi /2$, $\bar{\phi}=0$. Now we study the $\theta $ fluctuation near the
classical ground state, $A=\frac{\hbar ^2N}{8t}$ and $C=\frac{tN}2+\frac{%
\beta N^2}4=\frac N4(2t+\beta N)$. The effective action is

\begin{equation}
I\left( \theta _1\right) =\int_{\tau _i}^{\tau _f}\left[ \frac{\hbar ^2N}{8t}%
\dot{\theta}_1^2+\frac N4(2t+\beta N)\theta _1^2\right] d\tau .
\end{equation}
The motion of the fluctuation $N\theta _1$ is approximately a harmonic
oscillator with mass $\frac{\hbar ^2}{4tN}$ and frequency $\sqrt{\frac{%
2t(2t+\beta N)}{\hbar ^2}}$. Its characteristic length $\sigma _{N\theta }$
is determined by $\sigma _{N\theta }^2=N\sqrt{\frac{2t}{2t+\beta N}}$. Hence
the number fluctuation in one well is 
\begin{equation}
\Delta (N_L-N_R)=\left( N\sin \frac \pi 2\right) \Delta \theta =\sqrt{N\sqrt{%
\frac{2t}{2t+\beta N}}}.
\end{equation}
One can see that there is a singular point in $\beta N/2t=-1$, which
indicates a dividing behavior between coherent state-like and
Schr\"{o}dinger cat state-like ground state.

In the case of $\beta N/2t<-1$, the classical ground state is $\bar{\theta}%
=\theta _0$ or $\pi -\theta _0$, $\bar{\phi}=0$,where $\sin \theta
_0=2t/|\beta |N$. However, due to quantum tunneling between the two
classical ground states, the true ground state is an even-parity
Schr\"{o}dinger cat state. Inspecting the fluctuation near the classical
ground state, we find that $A=\frac{\hbar ^2N\sin \theta _0}{8t}=\frac{\hbar
^2}{4|\beta |}$ and $C=\frac{tN}{2\sin \theta _0}+\frac{\beta N^2}4\sin
^2\theta _0=\frac{|\beta |N^2}4\cos ^2\theta _0$. The effective action is

\begin{equation}
I\left( \theta _1\right) =\int_{\tau _i}^{\tau _f}\left[ \frac{\hbar ^2}{%
4|\beta |}\dot{\theta}_1^2+\frac{|\beta |N^2}4\cos ^2\theta _0\theta
_1^2\right] d\tau .
\end{equation}
The motion of the fluctuation $N\theta _1$ is approximately a harmonic
oscillator with mass $\frac{\hbar ^2}{2|\beta |N^2}$ and frequency $\frac{%
|\beta |N\cos \theta _0}\hbar $. Its characteristic length $\sigma _{N\theta
}$ is determined by $\sigma _{N\theta }^2=\frac{2N}{\cos \theta _0}$. Hence
the number fluctuation in one well is 
\begin{equation}
\Delta \left( N_L-N_R\right) =N\sin \theta _0\Delta \theta =\sin \theta _0%
\sqrt{\frac{2N}{\cos \theta _0}},
\end{equation}
which is different from the case of $\beta N/2t>-1$. This result is not the
true number fluctuation around the Schr\"{o}dinger cat state. It
indicates the fluctuation around the classical ground state. 
The true number fluctuation is of the 
order of $N^2$, referred to Ref.\cite{TLHYip} as a superfragmented state.

Finally we discuss the relative phase fluctuation. After performing the
Gaussian integration over $\theta _1$, the problem reduces to the
one-dimensional path integral with the effective action: 
\begin{equation}
I\left( \phi _1\right) =\int_{\tau _i}^{\tau _f}\left( A^{\prime }\dot{\phi}%
_1^2+C^{\prime }\phi _1^2\right) d\tau ,
\end{equation}
with 
\begin{eqnarray*}
A^{\prime } &=&\frac{\hbar ^2N^2\sin ^2\bar{\theta}}{8\left( E_{\theta
\theta }-\cot \bar{\theta}E_\theta \right) }, \\
C^{\prime } &=&\frac 12\left( E_{\phi \phi }-\frac{E_{\theta \phi }^2}{%
\left( E_{\theta \theta }-\cot \bar{\theta}E_\theta \right) }\right) \\
&&+i\frac d{d\tau }\left( \frac{\hbar N\sin \bar{\theta}E_{\theta \phi }}{%
4\left( E_{\theta \theta }-\cot \bar{\theta}E_\theta \right) }\right) .
\end{eqnarray*}

In the case of $\beta N/2t>-1$, the classical ground state is $\bar{\theta}%
=\pi /2$, $\bar{\phi}=0$. Now we inspect the $\phi $ fluctuation near the
classical ground state, $A^{\prime }=\frac{\hbar ^2N}{4\left( 2t+\beta
N\right) }$ and $C^{\prime }=\frac{tN}2$. The effective action is

\begin{equation}
I\left( \phi _1\right) =\int_{\tau _i}^{\tau _f}\left[ \frac{\hbar ^2N}{%
4\left( 2t+\beta N\right) }\dot{\phi}_1^2+\frac{tN}2\phi _1^2\right] d\tau .
\end{equation}
The motion of the fluctuation $\phi _1$ is approximately a harmonic
oscillator with mass $\frac{\hbar ^2N}{2\left( 2t+\beta N\right) }$ and
frequency $\sqrt{\frac{2t(2t+\beta N)}{\hbar ^2}}$. Its characteristic
length $\sigma _\phi $ is determined by $\sigma _\phi ^2=\frac 1N\sqrt{\frac{%
2t+\beta N}{2t}}$. Hence the relative phase fluctuation is 
\begin{equation}
\Delta \phi =\sqrt{\frac 1N\sqrt{\frac{2t+\beta N}{2t}}}.
\end{equation}

In the case of $\beta N/2t<-1$, the classical ground state is $\bar{\theta}%
=\theta _0$ or $\pi -\theta _0$, $\bar{\phi}=0$, where $\sin \theta
_0=2t/|\beta |N$. Inspecting the fluctuation near the classical ground
state, we find that $A^{\prime }=\frac{\hbar ^2\sin ^2\theta _0}{4|\beta
|\cos ^2\theta _0}$ and $C^{\prime }=|\beta |N^2\sin ^2\theta _0$. The
effective action is

\begin{equation}
I\left( \phi _1\right) =\int_{\tau _i}^{\tau _f}\left[ \frac{\hbar ^2\sin
^2\theta _0}{4|\beta |\cos ^2\theta _0}\dot{\phi}_1^2+|\beta |N^2\sin
^2\theta _0\phi _1^2\right] d\tau .
\end{equation}
The motion of the fluctuation $\phi _1$ is approximately a harmonic
oscillator with mass $\frac{\hbar ^2\sin ^2\theta _0}{2|\beta |\cos ^2\theta
_0}$ and frequency $\frac{2|\beta |N\cos \theta _0}\hbar $. Its
characteristic length $\sigma _\phi $ is determined by $\sigma _\phi ^2=%
\frac{\cos \theta _0}{2N\sin ^2\theta _0}$. Hence the relative phase
fluctuation is 
\begin{equation}
\Delta \phi =\sqrt{\frac{\cos \theta _0}{2N\sin ^2\theta _0}}.
\end{equation}

\section{Conclusion}

In summary, we study the quantum coherence phenomena in a double-well BEC
based on SU(2)-coherent-state path-integral. By this method we analytically
study the MQST phenomenon, the ground states of this system, and the existence
of macroscopic quantum superposition states. We find that MQST will happen in both
repulsive and attractive interaction cases, and analytically obtain the
dividing points $\beta N/2t=\pm 1$. When $\beta N/2t>-1$, both repulsive and
attractive Bose gases favor a coherent or a squeezed ground state, which is 
the coherent state for non-interaction case, the relative-number-squzeed state 
for replusive case, and the relative-phase-squzeed
state for the attractive case (see Eqs.(17) and (22)) respectively.
However when $\beta N/2t<-1$, attractive Bose gases favor a 
macroscopic quantum superposition state. 
The relative number fluctuation and relative phase fluctuation
between two-well condensates are obtained through path-integral technique.
For the attractive interaction case, the coherent gap of degenerate ground
states is obtained analytically with the help of the instanton technique.

It is noted that all the discussions in this paper does not impliy this kind of 
macroscopic quantum superposition states can be easily created. To inspect the decoherence,
one has to induce the interaction between thermal could and condensate. Dalvit 
et al. have studied the decoherence of a similar model in Ref \cite{Dalvit}, which is
beyond our two-mode approximation and semiclassical approach.
However, as an rough estimate, the high barrier and small splitting gap
imply high rate of decoherence. 
In fact, to the best of our knowledge, there is no experimental evidence of BEC cats up to now, 
although there are some proposals to create them\cite{TLH,Dalvit,cat}. Ref.\cite{Dalvit}
pointed out that such macroscopic quantum superposition states 
are extremely fragile to decoherence, and
suggested that the strategy of trap engineering and symmetrization of the environment
should be able to deal with that issue. 
The theoretical calculations performed in this paper can be extended to the
Bose gas in the non-symmetric double-well potential, and an effectively
two-component spinor condensate. Work along this line is still in progress.

We would like to thank Prof. Tin-Lun Ho and Prof. Qian Niu for stimulating
and helpful discussions during the Workshop and Conference on Modern Trends
of Condensed Matter Physics in Tsinghua University in Summer 2000. We would
like to acknowledge Yuan-Xiu Miao, Wei-Qiang Chen, and Hong-Yu Yang for
helpful discussions.


\begin{thebibliography}{99}
\bibitem{Rokshar}  D. Rokshar, cond-matt/9812260; R. W. Spekkens and J. E.
Spie, Phys. Rev. {\bf A59}, 3868 (1999).

\bibitem{Dalibard}  J. Javanainen and S. M. Yoo, Phys. Rev. Lett. {\bf 76},
161 (1996); Y. Castin and J. Dalibard, Phys. Rev. {\bf A55}, 4330 (1997); J.
I. Cirac, C. W. Gardiner, M. Naraschewski, and P. Zoller, Phys. Rev. {\bf A54%
} R3714 (1996).

\bibitem{TLH}  T. --L. Ho and C. V. Ciobanu, cond-matt/0011095.

\bibitem{MiIburn}  G. J. Milburn, J. F. Corney, E. M. Wright and D. F.
Walls, Phys. Rev. {\bf A55}, 4318(1997); J. F. Corney and G. J. Milburn,
Phys. Rev. {\bf A58}, 2399 (1998)

\bibitem{Vardi}  A. Vardi and J. R. Anglin, Phys. Rev. Lett. {\bf 86}, 568
(2001).

\bibitem{Parkins_Walls}  For a review, see A. S. Parkins and D. F. Walls,
Phys. Rep. {\bf 303}, 1 (1998); F. Dalfovo, S. Giorgini, L. P. Pitaevskii
and S. Stringari, Rev. Mod. Phys. {\bf 71}, 463 (1999); A. J. Leggett, Rev.
Mod. Phys. {\bf 73}, 307 (2001) and references therein.

\bibitem{Garg}  A. Garg and G. Kim, Phys. Rev. {\bf B45}, 12921 (1992); for
a review, see E. M. Chudnovsky and J. Tejada, {\it Macroscopic Quantum
Tunneling of the Magnetic Moment} (Cambridge University Press, Cambridge,
England, 1998).

\bibitem{Auerbach}  A. Auerbach, {\it Interacting Electrons and Quantum
Magnetism} (Springer, New York, 1994).

\bibitem{smerzi}  A. Smerzi, S. Fantoni, S. Giovanazzi, and S. R. Shenoy,
Phys. Rev. Lett. {\bf 79}, 4950 (1997); S. Raghavan, A. Smerzi, S. Fantoni
and S. R. Shenoy, Phys. Rev. {\bf A59}, 620 (1999) and references therein.

\bibitem{Coleman}  S. Coleman, {\it Aspect of Symmetry} (Cambridge
University Press, Cambridge, England, 1985), Chap. 7.

\bibitem{TLHYip}  T. --L. Ho and S. K. Yip, Phys. Rev. Lett. {\bf 84}, 4031(2000).

%\bibitem{Dalfovo}  F. Dalfovo and S. Stringari Phys. Rev. {\bf A53}, 2477
%(1996); R. J. Dodd, M. Edwards, C. J. Williams, C. W. Clark, M. J. Holland,
%P. A. Ruprecht and K. Burnett, Phys. Rev. {\bf A54}, 661 (1996).

%\bibitem{Bradley}  C. C. Bradley, C. A. Sackett, J. J. Tollett, and R. G.
%Hulet, Phys. Rev. Lett. {\bf 75}, 1687 (1995); C. C. Bradley, C. A. Sackett,
%and R. G. Hulet, Phys. Rev. Lett. {\bf 78}, 985 (1997).

\bibitem{Dalvit}  D. A. R. Dalvit, J. Dziarmaga, and W. H. Zurek, Phys. Rev. 
{\bf A62}, 013607 (2000) and references therein.

\bibitem{cat} J. I. Cirac, M. Lewenstein, K. M\o lmer, and P. Zoller, Phys. Rev. 
{\bf A57}, 1208 (1998); D. Gordon and C. M. Savage, Phys. Rev. {\bf A59}, 4623(1998). 


\end{thebibliography}
\end{document}